\documentclass[10pt]{article}
\usepackage{spconf,amsmath,graphicx,algorithm,amssymb, amsthm}
\usepackage{float}

\DeclareMathOperator*{\E}{\mathbb{E}}

\title{GRADED QUANTIZATION: DEMOCRACY FOR MULTIPLE DESCRIPTIONS IN COMPRESSED SENSING}

\name{Diego Valsesia \qquad Giulio Coluccia \qquad Enrico Magli \thanks{This work is supported by the European Research Council under the European Community's Seventh Framework Programme (FP7/2007-2013) / ERC Grant agreement n.279848.}}
\address{Politecnico di Torino (Italy) -- Dipartimento di Elettronica e Telecomunicazioni}

\begin{document}

\maketitle

\begin{abstract}

The compressed sensing paradigm allows to efficiently represent sparse signals by means of their linear measurements. However, the problem of transmitting these measurements to a receiver over a channel potentially prone to packet losses has received little attention so far. In this paper, we propose novel methods to generate multiple descriptions from compressed sensing measurements to increase the robustness over unreliable channels. In particular, we exploit the democracy property of compressive measurements to generate descriptions in a simple manner by partitioning the measurement vector and properly allocating bit-rate, outperforming classical methods like the multiple description scalar quantizer. In addition, we propose a modified version of the Basis Pursuit Denoising recovery procedure that is specifically tailored to the proposed methods. Experimental results show significant performance gains with respect to existing methods.

\end{abstract}


\vspace*{-0.1cm}
\section{INTRODUCTION}
\vspace*{-0.1cm}
\label{sec:intro}
In recent years, compressed sensing (CS) \cite{CS_donoho, candes2006compressive} has drawn great attention thanks to its remarkable results concerning signal recovery from vastly undersampled measurements. CS opened a new path to signal sampling and acquisition, showing that signals could be acquired directly in a compressed fashion, in the perspective of replacing the traditional approach based on collecting as many samples as possible and then removing the redundancy.

However, from the standpoint of practical systems, CS measurements typically need to be transmitted to a receiver. This raises the concern of how to protect the measurements when the communication channel is unreliable. A possible protection technique is represented by multiple description coding (MDC). MDC allows to increase the robustness to channel losses by creating multiple correlated representations of the original data, each carrying enough information to decode separately the data with a certain fidelity, in case of loss of the other descriptions. The decoding stage consists of \emph{side decoders}, able to recover a low-quality version of the data when a single description is received, and a \emph{central decoder}, able to jointly exploit all the descriptions for the best decoding quality. Multiple descriptions can be produced in various stages of the transmission chain. The MDC literature proposes methods which generate descriptions by preprocessing the source (\emph{e.g.}, a trivial method is to separate even and odd samples \cite{JayantMDC}), by applying a correlating transform \cite{PCT_Wang}, by using an ad-hoc quantizer \cite{mdsq}, or by applying channel codes to a layered version of the data to be transmitted \cite{UEP_Mohr}.

 In this paper we propose an MDC approach to CS to make it more robust to unreliable channels. Previous techniques are based on the idea of generating descriptions before sensing, \emph{e.g.}, in \cite{CSMDC} an image is partitioned into two subimages before sensing the wavelet coefficients of each. However, we argue that it could be much more appealing to create the descriptions \emph{after} the measurement process. This is supported by the fact that specialised hardware (\emph{e.g.}, \cite{SPCamera}) may be used to directly acquire the measurements, preventing any preprocessing of the signals. In particular, in this paper we propose two novel techniques, graded quantization (CS-GQ) and CS-SPLIT, for multiple description coding of the measurements. We compare their performance to CS-MDSQ, a system applying a multiple description scalar quantizer to the measurement vector. We will show that CS-GQ and CS-SPLIT, which exploit the democracy property of the measurements, have lower complexity and better performance than CS-MDSQ, which instead relies on a classic MDC method and, as such, does not fully exploit the properties of CS. Moreover, we show how the parameters of CS-GQ could be optimized on the expected description-loss probability. We also address the decoding process, proposing a variant of the Basis Pursuit Denoising (BPDN) algorithm for CS reconstruction, which is tailored to CS-GQ and can provide significant gains with respect to the standard BPDN. Finally, we provide a bound on the rate-distortion performance of CS-SPLIT and CS-MDSQ. 

\vspace*{-0.1cm}
\section{BACKGROUND AND NOTATION}
\vspace*{-0.1cm}
\label{sec:bkg}
CS is a novel theory for signal sensing and acquisition \cite{CS_donoho, candes2006compressive} able to acquire signals in an already compressed fashion, using fewer coefficients than dictated by the classical Nyquist-Shannon theory. Let us consider a signal $\mathbf{x} \in \mathbb{R}^n$, having a sparse representation under basis $\Psi \in \mathbb{R}^{n \times n}$:
$\mathbf{x} = \Psi\boldsymbol{\theta} \hspace{0.2cm} \mathrm{with} \hspace{0.2cm} \left\Vert \boldsymbol{\theta} \right\Vert_0 = k \ll n$, being $\left\Vert \boldsymbol{\theta} \right\Vert_0$ the $l_0$ norm of $\boldsymbol{\theta}$, \emph{i.e.}, the number of its nonzero entries.
We acquire measurements as a vector of random projections $\mathbf{y} = \Phi\mathbf{x} = \Phi\Psi\boldsymbol{\theta}$, $\mathbf{y} \in \mathbb{R}^m$, using a sensing matrix $\Phi \in \mathbb{R}^{m \times n}$.
A very popular way to recover the original signal from the measurements is to solve an optimization problem that minimises the $l_0$ norm of the signal in the domain where the signal is sparse. However, this problem is computationally intractable due to its NP-hard complexity, so it is common to consider a relaxed form using the $l_1$ norm, which can be solved by means of convex optimization techniques. In presence of bounded noise it is common to consider an $l_2$ norm constraint using a bound $\epsilon$ on the noise norm, see \eqref{BPDN}. This is also used when dealing with quantization, which is an important issue in practical systems that require finite precision in the representation of the measurements. In the remainder of the paper we deal with quantized measurements.
\vspace*{-0.05cm}
\begin{align}
\label{BPDN}
\hat{\boldsymbol{\theta}} &= \arg \underset{\boldsymbol{\theta}}{\text{min}} \left\Vert \boldsymbol{\theta} \right\Vert_1 \hspace{.2cm} \text{subject to} \hspace{.2cm} \left\Vert \mathbf{y} - \Phi\Psi\boldsymbol{\theta} \right\Vert_2 \leq \epsilon 
\end{align}  
These methods are successful provided that enough measurements have been acquired, typically $m = O\left( k\log \frac{n}{k} \right)$.
It is relevant to notice the \emph{democracy} of the measurements \cite{Democracy_Laska}, in the sense that each contributes roughly in the same manner to the reconstruction of the signal and no measurement carries significantly more information than the others.

\vspace*{-0.1cm}
\section{MULTIPLE DESCRIPTIONS FOR CS}
\vspace*{-0.1cm}
\begin{figure}[t]
  \centering
  \centerline{\includegraphics[width=0.7\linewidth]{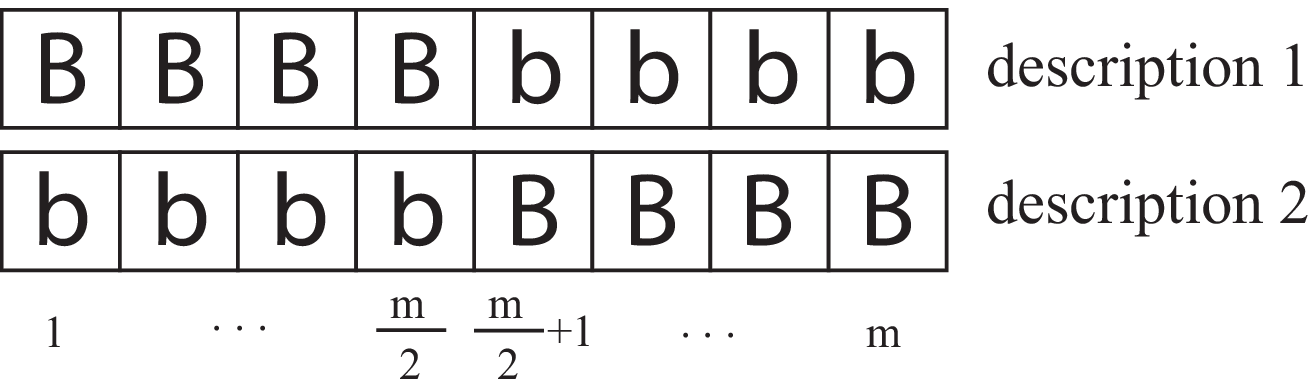}}
%  \vspace{2.0cm}
\medskip
\vspace*{-0.55cm}
\caption{\small{CS-GQ}}
\vspace*{-0.5cm}
\label{graded}
\end{figure}

In this section we describe the proposed multiple description techniques with particular focus on the case of two balanced descriptions. CS-GQ and CS-SPLIT rely on partitioning the vector of measurements and quantize the subsets. Those methods are compared against CS-MDSQ that is derived from the classical MDC technique of the MDSQ. The measurements are quantized with a uniform scalar quantizer (with the exception of CS-MDSQ). More complex quantizers could also be used, \emph{e.g.}, the Lloyd-Max method or vector quantization, but they are regarded as computationally too complex and with little or no gain as shown in \cite{Milenkovic}. 

 \vspace*{-0.25cm}
\subsection{Graded Quantization (CS-GQ)}
 \vspace*{-0.05cm}
\subsubsection{Encoding}
CS-GQ (see Fig. \ref{graded}) creates two descriptions by having each measurement coded in a redundant way. The first description contains the first $\frac{m}{2}$ measurements quantized with $2^B$ levels and the other half with $2^b$ levels, with $B \geq b$, corresponding to quantization step sizes $\Delta_1$ and $\Delta_2$ respectively. Conversely, the second description uses $2^b$ and $2^B$ levels respectively. Since the uniform scalar quantizer produces an embedded codebook, it is possible to regard the measurement quantized with the lower rate as being made of the most significant bits of the high-rate version of the same measurement. Thanks to the democracy of the measurements, it is indifferent which measurements are actually finely or coarsely quantized. This is also the reason why the two descriptions turn out to be balanced. It would also be possible to obtain unbalanced descriptions by varying the ratio of measurements quantized at high and low rates; this is left for further work.
 
 \vspace*{-0.14cm}
\subsubsection{Decoding}
 \vspace*{-0.05cm}
Because each description contains as many measurements as acquired during the sensing process, any of them can be decoded separately to provide a basic quality level, essentially depending on the quantization step sizes employed (see Sec. \ref{sec:exp_perf}). Since there are two distinct groups of measurements (fine and coarse quantization) inside each description, we can use this knowledge to improve the BPDN algorithm in the reconstruction phase at the side decoders. In particular we set two $l_2$ constraints, one for each subset, instead of a single one. Moreover, we add two extra constraints called \emph{quantization consistency} to ensure that the measurements of the reconstructed signal fall inside the quantization bins of size $\Delta_1$ and $\Delta_2$ of the original measurements. Hence, reconstruction at each side decoder is performed solving the following problem:

\vspace*{-0.7cm}
\begin{align*}
\hspace*{0.015cm}\hat{\boldsymbol{\theta}}=\arg\underset{\boldsymbol{\theta}}{\text{min}}\left\Vert \boldsymbol{\theta}\right\Vert _{1}\hspace{0.15cm}\text{subject to}\hspace{0.05cm}\begin{cases}
\left\Vert \mathbf{y}^{(1)}-\Phi^{(1)}\Psi\boldsymbol{\theta}\right\Vert _{2}&\leq\epsilon_{1}\\
\left\Vert \mathbf{y}^{(1)}-\Phi^{(1)}\Psi\boldsymbol{\theta}\right\Vert _{\infty}&\leq\frac{\Delta_{1}}{2}\\
\left\Vert \mathbf{y}^{(2)}-\Phi^{(2)}\Psi\boldsymbol{\theta}\right\Vert _{2}&\leq\epsilon_{2}\\
\left\Vert \mathbf{y}^{(2)}-\Phi^{(2)}\Psi\boldsymbol{\theta}\right\Vert _{\infty}&\leq\frac{\Delta_{2}}{2}
\end{cases}
\end{align*}

\begin{figure}[t]
  \centering
  \centerline{\includegraphics[width=0.7\linewidth]{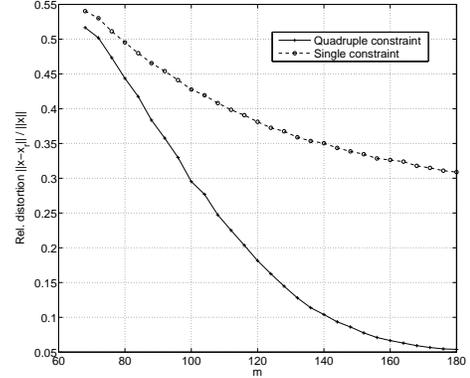}}
%  \vspace{2.0cm}
\medskip
\vspace*{-0.55cm}
\caption{\small{Performance of modified BPDN for side decoding of CS-GQ vs. standard BPDN}}
\vspace*{-0.4cm}
\label{4C-BPDN}
\end{figure}  
 
In our simulations (see Fig. \ref{4C-BPDN}) the modified reconstruction problem shows significant gains with respect to standard BPDN \eqref{BPDN}. This is mainly due to the double $l_2$ constraint, while quantization consistency provides a small gain overall. The central decoder instead always selects the measurement with finer quantization and runs the standard BPDN procedure \eqref{BPDN}. This means that some pieces of information are discarded by the central decoder because they are redundant. The issue of redundancy is central in MDC. The descriptions must share some information on the signal, in order to be independently decodable. In the proposed CS-GQ scheme, the amount of redundancy can be tuned in a very flexible way through the choice of parameters $b$ and $B$, depending on the desired level of quality at the side and central decoders. These levels may also depend on the channel error or packet loss rate, in that frequent losses are typically coped with by selecting a higher degree of redundancy.

\begin{figure}[t]
  \centering
  \centerline{\includegraphics[width=0.7\linewidth]{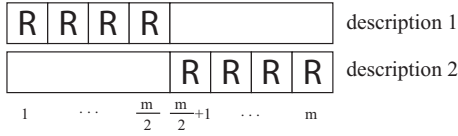}}
%  \vspace{2.0cm}
\medskip
\vspace*{-0.55cm}
\caption{\small{CS-SPLIT}}
\vspace*{-0.65cm}
\label{split}
\end{figure} 
 
 \vspace*{-0.3cm}
\subsection{CS-SPLIT}
\vspace*{-0.15cm}
CS-SPLIT (see Fig. \ref{split}) consists in quantizing all the measurements with rate $R$ and partitioning the measurements vector into two subsets with $\frac{m}{2}$ measurements each. The side decoders receive only one subset and recover the signal using \eqref{BPDN}, where $\Phi$ is the appropriate submatrix of the original sensing matrix. Instead, the central decoder can use all the measurements. CS-SPLIT can be regarded as a special case of CS-GQ in which $B=R$ and $b=0$. In this special case no redundant information is transmitted. In fact, the central decoder does not discard any information as it happens in CS-GQ. CS-SPLIT may be an appealing solution thanks to its extreme simplicity, as it does not require any additional processing other than splitting the measurement vector. However, we shall discuss in Sec. \ref{sec:exp_perf} how CS-SPLIT always outperforms CS-GQ when the number of measurements is high. 

\vspace*{-0.3cm}
\subsection{CS-MDSQ}
\vspace*{-0.15cm}
CS-MDSQ creates multiple descriptions of the measurements using a special quantizer called MDSQ. The MDSQ is a general technique for MDC developed before the advent of CS, so it does not leverage the democracy property as the previous systems do. The MDSQ can be optimised at several levels and can be tuned to operate at different points on the central distortion vs. side distortion curve. For the results in this paper we used the optimization method outlined in \cite{mdsq}, using nested assignment for the index assignment matrix. 

\vspace*{-0.1cm}
\section{Rate-Distortion performance for CS-SPLIT and CS-MDSQ}
\vspace*{-0.1cm}
\label{sec:RD}
\newtheorem{thm}{Theorem}
\begin{thm}
Consider a $k$-sparse signal $\mathbf{x}\in\mathbb{R}^{n}$ and its measurements $\mathbf{y}=\Phi \mathbf{x}$, $\mathbf{y}\in \mathbb{R}^m$. Assume that the nonzero entries have zero mean and variance $\sigma_x^2$. Assume that the entries of the sensing matrix $\Phi$ are i.i.d. Gaussian random variables with $\Phi_{ij}\sim\mathcal{N}(0,\frac{1}{m})$ and such that $m > 60 \log{n} $ and $k<\frac{1}{4}(\frac{1}{\mu}+1)$, where $\mu$ is the coherence of $\Phi$. Furthermore, assume that the BPDN algorithm is used for reconstruction.  Then the distortion $D=\left\Vert \mathbf{x}-\mathbf{\hat{x}} \right\Vert_2^2$ in the reconstructed signal as a function of rate $R$ is bounded as follows, with high probability:
\begin{equation}
\vspace*{-0.35cm}
\label{thm_side}
\frac{k^{2}}{m}\sigma_x^2 2^{-2R} \leq D_{\mathrm{side}}(R) \leq \frac{4k\sigma_x^2 2^{-2R}}{1-\sqrt{\frac{30\log n}{m}}(4k-1)} 
\end{equation}
\vspace*{-0.05cm}
\begin{equation}
\label{thm_central}
\frac{k^{2}}{m}\sigma_x^2 2^{-2R} \leq D_{\mathrm{central}}(R) \leq \frac{4k\sigma_x^2 2^{-2R}}{1-\sqrt{\frac{15\log n}{m}}(4k-1)} 
\end{equation}
\end{thm}

A very similar argument can be used to analyse CS-MDSQ. The assumptions are the same as before, but the MDSQ performance is limited by the Ozarow bound \cite{ozarow}.
\begin{thm}
\vspace*{-0.15cm}
 Under the same hypotheses of Theorem 1, the distortion $D=\left\Vert \mathbf{x}-\mathbf{\hat{x}} \right\Vert_2^2$ in the reconstructed signal as a function of rate $R$ is bounded as follows, with high probability:
 \vspace*{-0.45cm}
\begin{equation}
\frac{\sigma_x^2}{m} k^2 2^{-2R} \leq D_\mathrm{side}(R) \leq \frac{4\sigma_x^2 k 2^{-2R}}{1-\sqrt{\frac{15\log n}{m}}\left(4k-1\right)}
\end{equation}
\vspace*{-0.4cm}
\begin{equation}
\label{cs-mdsq_ctrl}
\frac{\sigma_x^2}{m} k^2 2^{-4R}\gamma_D \leq D_\mathrm{central}(R) \leq \frac{4\sigma_x^2k 2^{-4R}\gamma_D}{1-\sqrt{\frac{15\log n}{m}}\left(4k-1\right)}
\end{equation}
with 
\vspace*{-0.5cm}
\small

\begin{equation*}
\gamma_{D}=
\left[ 1-\left(\left(1-\frac{D_{\mathrm{sm,side}}}{\frac{\sigma_x^2}{m} k}\right)-\sqrt{\frac{D_{\mathrm{sm,side}}^2}{\frac{\sigma_x^4}{m^2} k^2}-2^{-4R}}\right)^{2} \right]^{-1}
\end{equation*}
\normalsize
and $D_{\mathrm{sm,side}} =  \E \left[ \left( y_i - \mathcal{Q}\left(y_i\right) \right)^2 \right]$. 
\end{thm}

The proofs of the previous results and experimental results showing their validity are omitted for brevity. By looking at the lower bounds it can be seen that CS-SPLIT can potentially achieve $( D_\mathrm{side},D_\mathrm{central})$ points that are unavailable for CS-MDSQ.

\vspace*{-0.1cm}
\section{EXPERIMENTAL RESULTS}
\vspace*{-0.1cm}
\label{sec:exp_perf}

\begin{figure}[t]

\begin{minipage}[b]{.497\linewidth}
  \centering
  \centerline{\includegraphics[width=0.97\linewidth]{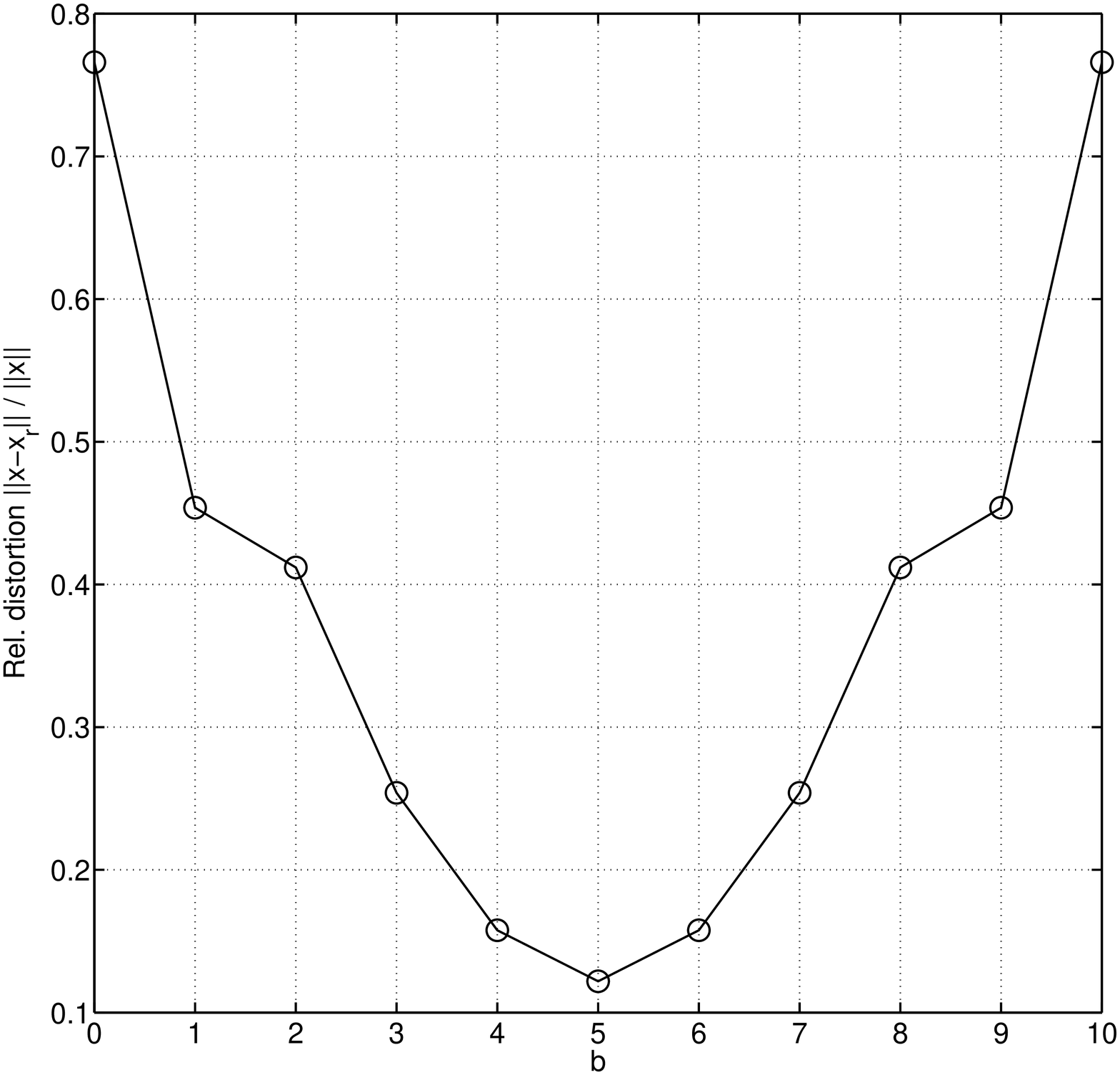}}
%  \vspace{1.5cm}
  \centerline{(a) Side relative distortion}\medskip
\end{minipage}
\hfill
\begin{minipage}[b]{0.495\linewidth}
  \centering
  \centerline{\includegraphics[width=0.99\linewidth]{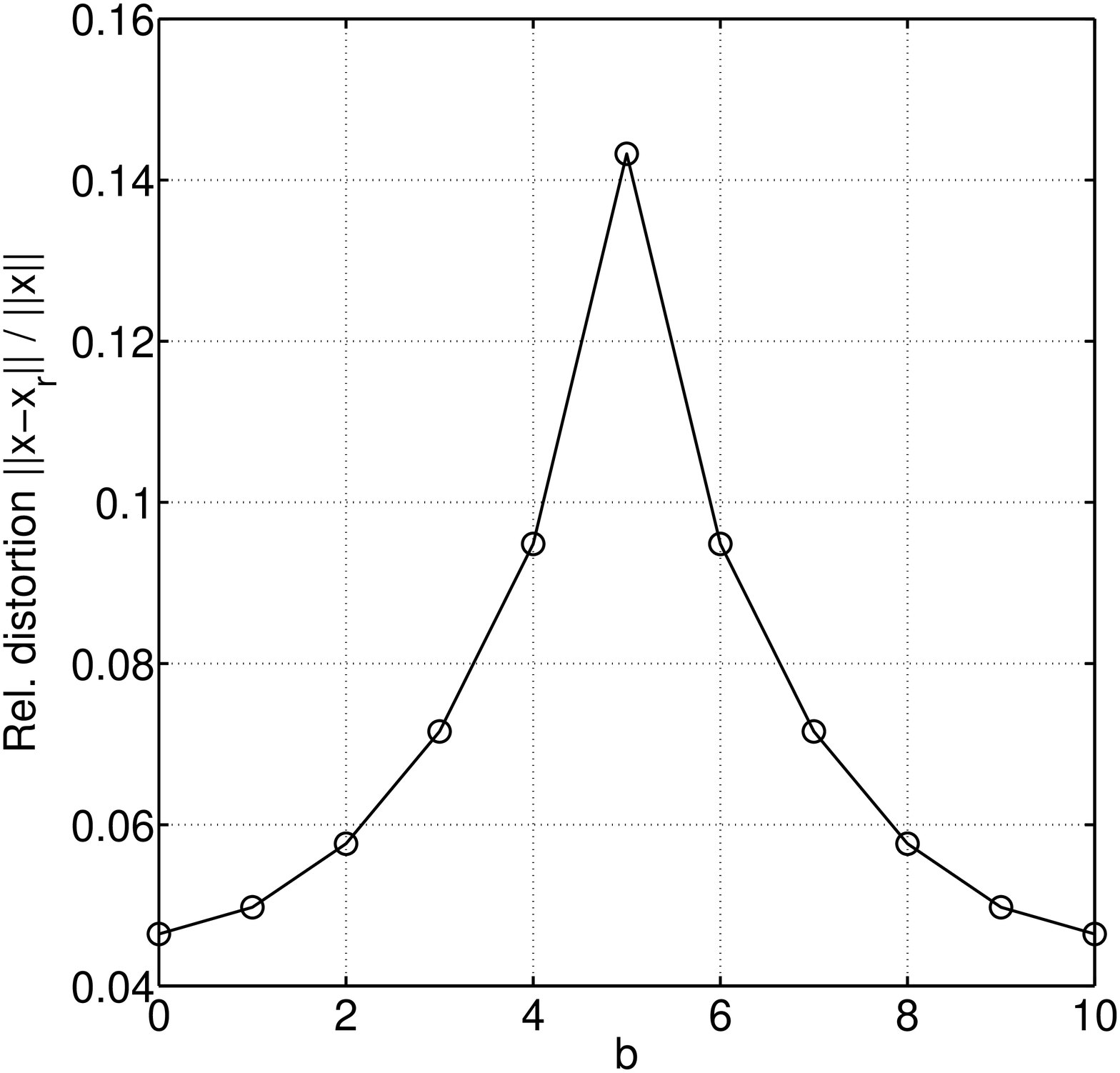}}
%  \vspace{1.5cm}
  \centerline{(b) Central relative distortion}\medskip
\end{minipage}
\vspace*{-0.8cm}
\small{\caption{$n=256$, $k=10$, $m=50$, Gaussian sensing matrix}}
\label{perf_vs_redundancy}
\vspace*{-0.4cm}
\end{figure}

 \begin{figure}
  \centering
  \centerline{\includegraphics[width=0.65\linewidth]{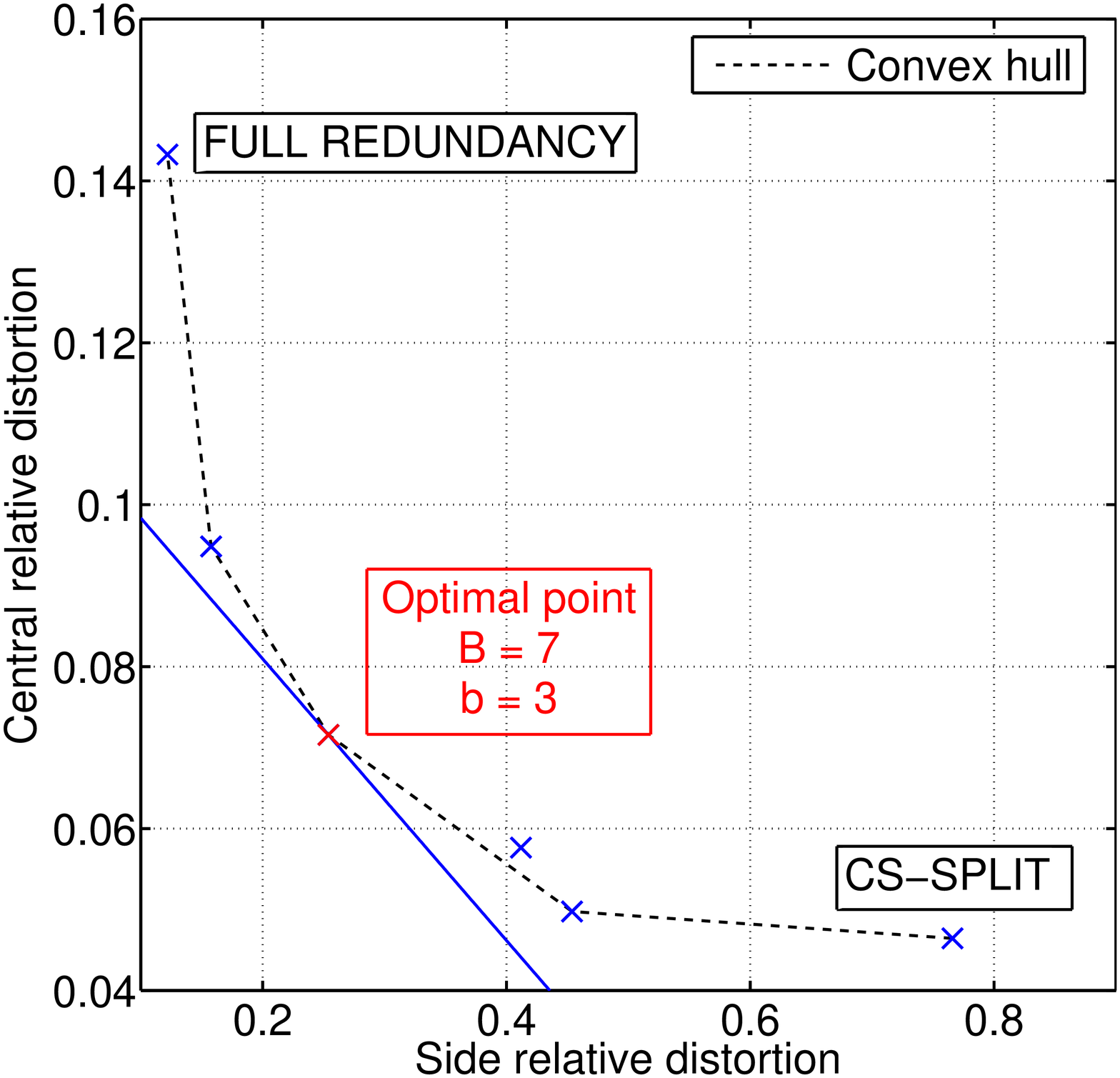}}
%  \vspace{2.0cm}
\medskip
\vspace*{-0.5cm}
\small{\caption{Central vs. side distortion tradeoff plot. $n=256$, $k=10$, $m=50$, Gaussian sensing matrix. Optimal point for $p=0.08$.}}
\vspace*{-0.2cm}
\label{centr_side_tradeoff}
\end{figure}

\begin{figure}[t]

%\vspace*{-0.65cm}

\begin{minipage}[h]{.495\linewidth}
  \centering
  \centerline{\includegraphics[width=0.998\linewidth]{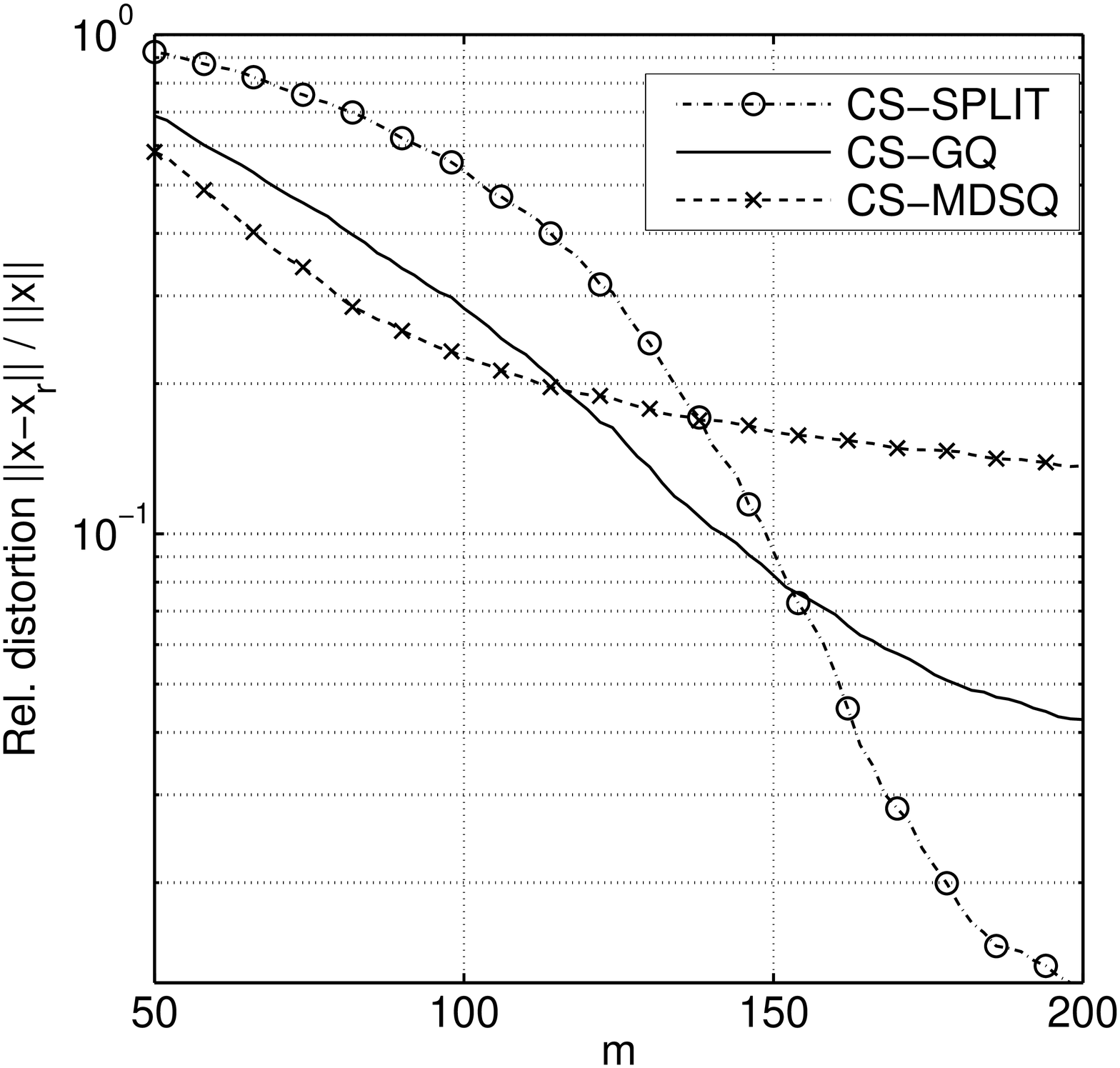}}
%  \vspace{1.5cm}
  \centerline{(a) Side relative distortion}\medskip
\end{minipage}
\hfill
\begin{minipage}[h]{0.495\linewidth}
  \centering
  \centerline{\includegraphics[width=0.980\linewidth]{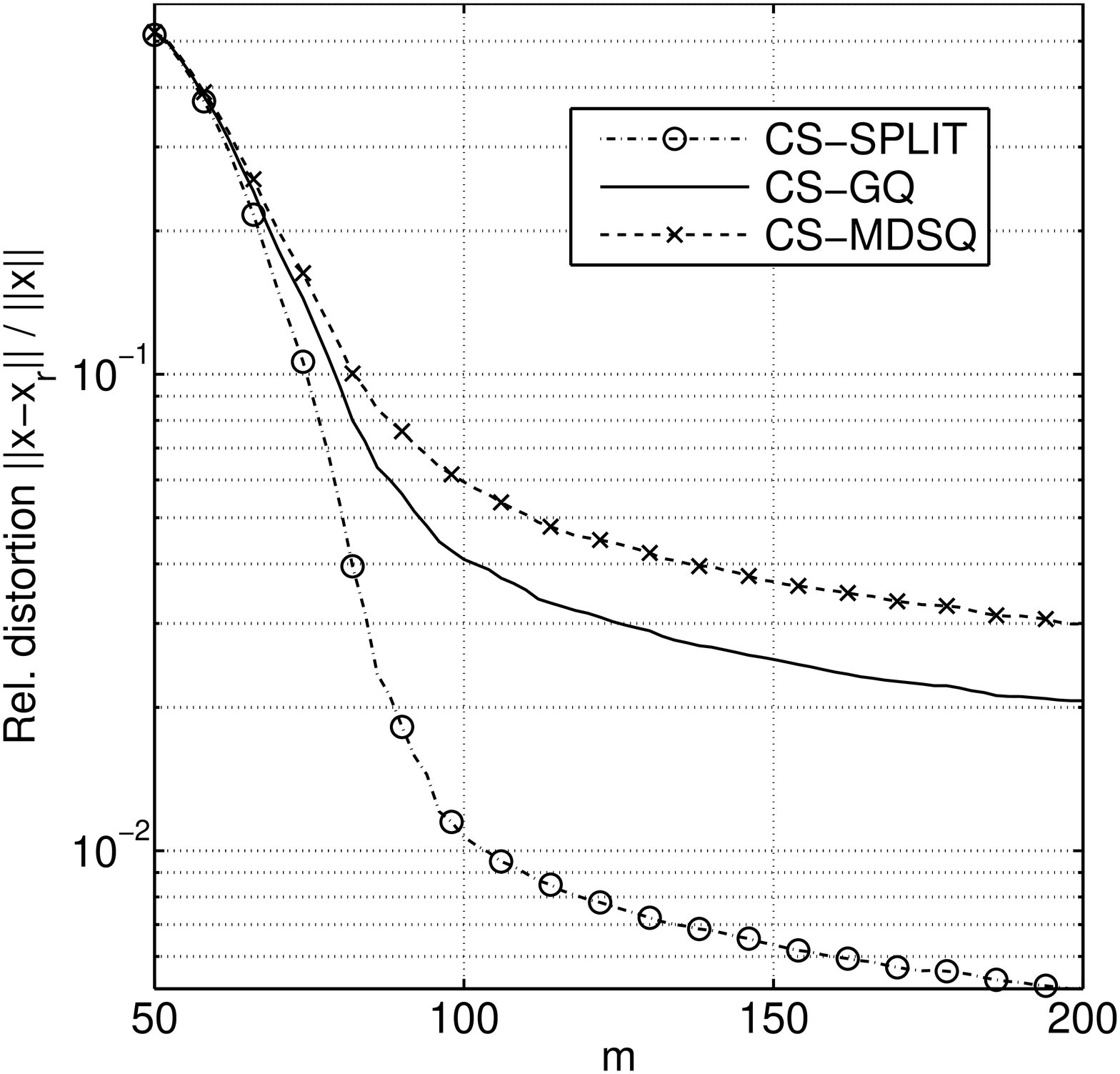}}
%  \vspace{1.5cm}
  \centerline{(b) Central relative distortion}\medskip
\end{minipage}
\vspace*{-0.4cm}
\small{\caption{$n=256$, $k=20$. CS-GQ: $B=6$, $b=2$. CS-SPLIT: $R=8$. CS-MDSQ: $R=4$. Gaussian sensing matrix}}
\label{perf_vs_m}
\vspace*{-0.5cm}
\end{figure}

\begin{figure}[t]
%\vspace*{-0.1cm}
\begin{minipage}[h]{.495\linewidth}
  \centering
  \centerline{\includegraphics[width=0.998\linewidth]{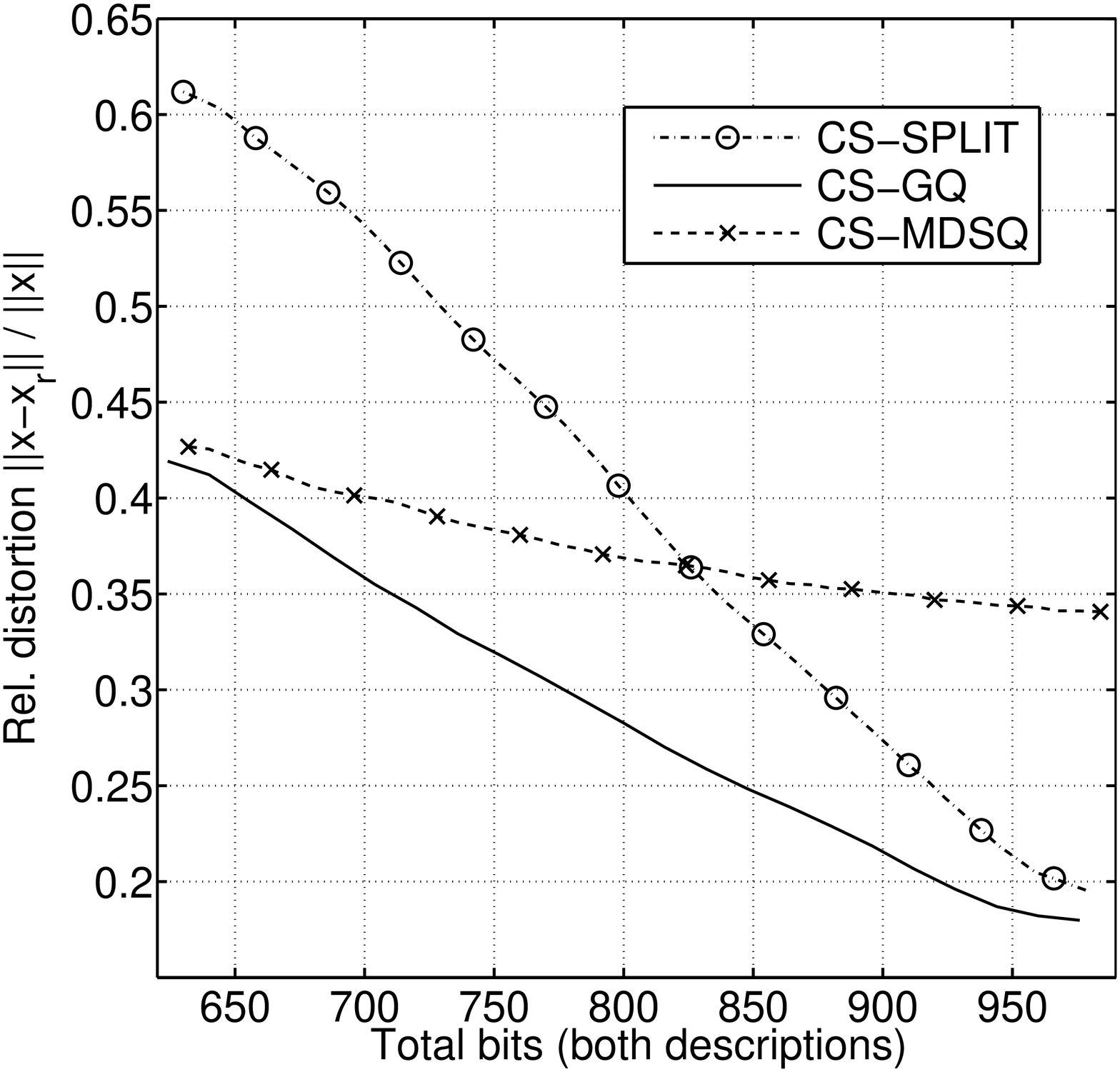}}
%  \vspace{1.5cm}
  \centerline{(a) Side relative distortion}\medskip
\end{minipage}
\hfill
\begin{minipage}[h]{0.495\linewidth}
  \centering
  \centerline{\includegraphics[width=0.980\linewidth]{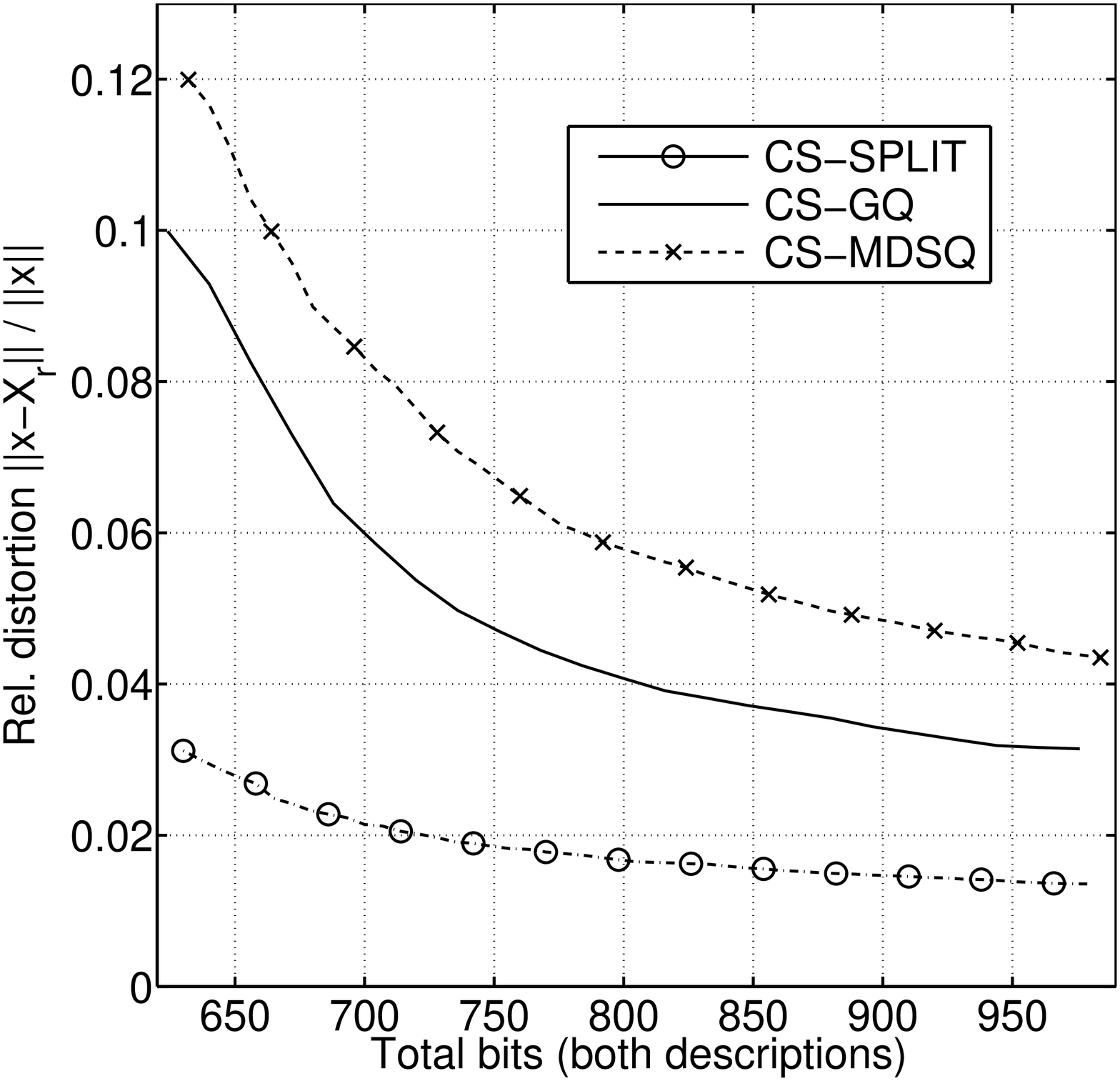}}
%  \vspace{1.5cm}
  \centerline{(b) Central relative distortion}\medskip
\end{minipage}
\vspace*{-0.4cm}
\small{\caption{$n=256$, $k=20$. CS-GQ: $m \in \left[78,122\right]$, $B=6$, $b=2$. CS-SPLIT: $m \in \left[90,140\right]$, $R=7$. CS-MDSQ: $m \in \left[79,123\right]$, $R=4$. Gaussian sensing matrix}}
\vspace*{-0.45cm}
\label{perf_vs_bits}
\end{figure}

In this section we compare the reconstruction performance of the side and central decoders of the proposed methods. As distortion we consider the normalised error norm $\frac{\left\Vert \mathbf{x}-\mathbf{\hat{x}} \right\Vert_2}{\left\Vert \mathbf{x} \right\Vert_2}$.
First, we characterise the relative performance of CS-GQ and CS-SPLIT at equal bit-rate for the same number of measurements. Suppose that CS-SPLIT uses a rate of $R$ bits per measurement, then we must have $B+b=R$. From Fig. \ref{perf_vs_redundancy} we can see that CS-GQ improves side decoding performance for increasing redundancy until the full redundancy case ($B=b$) is reached, while at the same time central decoding performance worsens. CS-SPLIT is the extreme case with best central but worst side performance. The appropriate values of $B$ and $b$ can be selected from the trade-off plot show in Fig. \ref{centr_side_tradeoff}, according to the desired trade-off between side and central distortion. If a memoryless channel has a probability $p$ of losing a description, we can define an average distortion $D_\mathrm{avg} = 2 \cdot D_\mathrm{side} \cdot p(1-p) + D_\mathrm{cent} \cdot(1-p)^2 + p^2$. Keeping in mind that $D_\mathrm{cent}$ is a function of $D_\mathrm{side}$ and letting $\frac{\mathrm{d}D_\mathrm{avg}}{\mathrm{d}D_\mathrm{side}} = 0$, we get $\frac{\mathrm{d}D_\mathrm{cent}}{\mathrm{d}D_\mathrm{side}} = -\frac{2p}{1-p}$. This is the slope of the straight line that is tangent to the trade-off plot in the point representing the optimal trade-off between central and side distortion, so it can be used to determine the optimal values of $B$ and $b$ for a channel with given packet-loss rate. In our case the feasible points are a discrete set and this method can only select one of the points lying on the convex hull. Notice that $p=0$ selects CS-SPLIT (there are no losses, hence we seek the best central performance and minimum redundancy) and that for $p \rightarrow 1$ full redundancy is approached where CS-GQ behaves like a repetition code.
 
 We should also notice that CS-GQ providing gains at the side decoders, with respect to CS-SPLIT, is the typical behaviour when $m$ is small (\emph{e.g.}, $m$ is so small that CS-SPLIT fails side decoding, or bigger but still in the regime in which extra measurements are more important than finer quantization). In fact, when $m$ grows very large, CS-SPLIT is always favourable due to the convenience in investing the budgeted bits in finer quantization rather than in extra measurements. 
 
 Comparing the performance with respect to CS-MDSQ, we can see that if the number of measurements is fixed a priori, CS-GQ and CS-SPLIT can benefit from higher quantization rates and thus outperform CS-MDSQ most of the times as shown in Fig. \ref{perf_vs_m}. CS-MDSQ can be advantageous only when we are forced to acquire few measurements, but for the same total bit-rate a slight oversampling can allow to use the more efficient graded quantization. Fig. \ref{perf_vs_bits} shows the case of a fully tunable system in which both the number of measurements and the rate can be adjusted. Also in this case we can see that graded quantization has lower reconstruction distortion both for central and side decoding in many practical settings. 

\vspace*{-0.2cm}
\section{CONCLUSIONS}
\vspace*{-0.1cm}
In this paper we showed how the democracy property of CS measurements enables to address the multiple descriptions problem in a simple and yet effective manner. We proposed methods to generate multiple descriptions from CS measurements, without the need of preprocessing the signal. As a term of comparison with classical literature on MDC, CS-MDSQ is derived from the MDSQ, and does not explicitly rely on properties of CS. In fact, we showed that it can be outperformed by the other proposed methods in many cases. CS-GQ and its limit case CS-SPLIT leverage the democracy of the measurements to create balanced descriptions in a straightforward manner, yet allowing great flexibility in selecting the desired trade-off between central and side distortion. 

\clearpage
\pagebreak


\begin{thebibliography}{10}

\bibitem{CS_donoho}
D.L. Donoho,
\newblock ``Compressed sensing,''
\newblock {\em Information Theory, IEEE Transactions on}, vol. 52, no. 4, pp.
  1289 --1306, april 2006.

\bibitem{candes2006compressive}
E.J. Cand{\`e}s,
\newblock ``Compressive sampling,''
\newblock in {\em Proceedings oh the International Congress of Mathematicians:
  Madrid, August 22-30, 2006: invited lectures}, 2006, pp. 1433--1452.

\bibitem{JayantMDC}
N.~S. Jayant,
\newblock ``Subsampling of a {DPCM} speech channel to provide two
  ``self-contained'' half-rate channels,''
\newblock {\em Bell System Technical Journal}, vol. 60, no. 4, pp. 501--509,
  Apr. 1981.

\bibitem{PCT_Wang}
Y.~Wang, M.T. Orchard, V.~Vaishampayan, and A.R. Reibman,
\newblock ``Multiple description coding using pairwise correlating
  transforms,''
\newblock {\em Image Processing, IEEE Transactions on}, vol. 10, no. 3, pp. 351
  --366, mar 2001.

\bibitem{mdsq}
V.A. Vaishampayan,
\newblock ``Design of multiple description scalar quantizers,''
\newblock {\em IEEE Transactions on Information Theory}, vol. 39, no. 3, pp.
  821--834, May 1993.

\bibitem{UEP_Mohr}
A.E. Mohr, E.A. Riskin, and R.E. Ladner,
\newblock ``Unequal loss protection: graceful degradation of image quality over
  packet erasure channels through forward error correction,''
\newblock {\em Selected Areas in Communications, IEEE Journal on}, vol. 18, no.
  6, pp. 819 --828, june 2000.

\bibitem{CSMDC}
D.~Liu, D.~Gao, and G.~Shi,
\newblock ``Compressive sensing-based multiple description image coding for
  wireless channel,''
\newblock in {\em Wireless Communications and Signal Processing (WCSP), 2010
  International Conference on}, oct. 2010, pp. 1 --5.

\bibitem{SPCamera}
M.F. Duarte, M.A. Davenport, D.~Takhar, J.N. Laska, T.~Sun, K.F. Kelly, and
  R.G. Baraniuk,
\newblock ``Single-pixel imaging via compressive sampling,''
\newblock {\em IEEE Signal Processing Magazine}, vol. 25, no. 2, pp. 83--91,
  March 2008.

\bibitem{Democracy_Laska}
J.N. Laska, P.T. Boufounos, M.A. Davenport, and R.G. Baraniuk,
\newblock ``Democracy in action: Quantization, saturation, and compressive
  sensing,''
\newblock {\em Applied and Computational Harmonic Analysis}, vol. 31, no. 3,
  pp. 429 -- 443, 2011.

\bibitem{Milenkovic}
W.~Dai, H.V. Pham, and O.~Milenkovic,
\newblock ``Information theoretical and algorithmic approaches to quantized
  compressive sensing,''
\newblock {\em IEEE Transactions on Information Theory}, vol. 59, no. 7, pp.
  1857--1866, July 2011.

\bibitem{ozarow}
L.~Ozarow,
\newblock ``On a source coding problem with two channels and three receivers,''
\newblock {\em Bell System Technical Journal}, vol. 59, no. 10, pp. 1909--1921,
  December 1980.

\end{thebibliography}
\end{document}